# Thermodynamics of amide + amine mixtures. 5. Excess molar enthalpies of *N,N*-dimethylformamide or *N,N*-dimethylacetamide + *N*-propylpropan-1-amine, + *N*-butylbutan-1-amine, + butan-1-amine, or + hexan-1-amine systems at 298.15 K. Application of the ERAS model

Fernando Hevia[1], Karine Ballerat-Busserolles[2], Yohann Coulier[2], Jean-Yves Coxam[2], Juan Antonio González[1]*, Isaías García de la Fuente[1], José Carlos Cobos[1]

[1] G.E.T.E.F., Departamento de Física Aplicada, Facultad de Ciencias, Universidad de Valladolid. Paseo de Belén, 7, 47011 Valladolid, Spain.

[2] Institut de Chimie de Clermont Ferrand, University Clermont Auvergne, CNRS UMR 6296, SIGMA Clermont, F-63000 Clermont-Ferrand, France.

*e-mail: jagl@termo.uva.es; Tel: +34-983-423757

# Abstract

Excess molar enthalpies, $H_m^E$, over the whole composition range have been determined for the liquid mixtures *N,N*-dimethylformamide (DMF) or *N,N*-dimethylacetamide (DMA) + butan-1-amine (BA), or + hexan-1-amine (HxA), or + *N*-propylpropan-1-amine (DPA), or *N*-butylbutan-1-amine (DBA) at 298.15 K and at 0.1 MPa using a BT2.15 calorimeter from Setaram adapted to work in dynamic mode at constant temperature and pressure. All the $H_m^E$ values are positive, indicating that interactions between like molecules are predominant. The replacement of DMF by DMA in systems with a given amine leads to lower $H_m^E$ results, which have been ascribed to stronger amide-amide interactions in DMF mixtures. The replacement of HxA by DPA in systems with a given amide leads to slightly higher $H_m^E$ values, as interactions between unlike molecules are weaker for the latter. Structural effects in the investigated solutions are also present, since the corresponding excess molar volumes ($V_m^E$), previously determined, are negative or slightly positive. The systems have been characterized in terms of the ERAS model reporting the interaction parameters. The model correctly describes both $H_m^E$ and $V_m^E$. The application of the model suggests that, in the systems under study, solvation effects are of minor importance and that physical interactions are dominant.

# 1. Introduction

It is well-known that a suitable approach for the investigation of the highly complex chemical environment of proteins is to study small organic molecules whose functional groups are similar to those present in the biomolecule [1]. The systematic physical and chemical characterization of such molecules and of their mixtures in terms of thermodynamic, transport and dielectric properties is necessary in this framework. The study of amide + amine systems is relevant, as it allows to gain insight into the behavior of the amide group when it is surrounded by different environments. In fact, the hydrogen-bonded structures where the amide group is involved can show very different biological activities depending on the mentioned environments [2]. On the other hand, the strong polarity of amides, which in the case of tertiary amides leads to the creation of a certain local order [3, 4], together with their high solvating capability and liquid state range –due to their ability to form hydrogen bonds– [5], makes them a very important kind of organic solvents. Similarly, amines are also an important class of substances since many biological relevant molecules contain the amine group [6-8]. In addition, the low vapor pressure of amines makes them useful in green chemistry. Thus, mixtures containing amines are being investigated to be used in $CO_2$ capture [9] and, interestingly, many of the ions of the technically important ionic liquids are related to amine groups [10].

In previous works, we have measured densities, speeds of sound and refractive indices of *N,N*-dimethylformamide (DMF) [11], or *N,N*-dimethylacetamide (DMA) [12] + *N*-propylpropan-1-amine (DPA) or + butan-1-amine (BA) at (293.15-303.15) K, and + *N*-butylbutan-1-amine (DBA) or + hexan-1-amine (HxA) at 298.15 K. In addition, we have reported low-frequency permittivity measurements of the mentioned systems and of the DMF + aniline mixture at (293.15-303.15) K [13, 14]. This database has been interpreted in terms of solute-solvent interactions and structural effects. We have also applied the ERAS [15] and the Kirkwood-Fröhlich models [16-19] to the study of amine + amide mixtures. The latter is useful for the calculation of the Balankina relative excess Kirkwood correlation factors [20], which provide information on the dipole correlations present in the considered systems. Calorimetric data are essential for the study of the type and strength of interactions present in liquid mixtures. As the data available in the literature on excess molar enthalpies, $H_m^E$, for amine + amide mixtures is scarce [21-23], we continue this series of works reporting $H_m^E$ values for DMF or DMA + DPA, or + DBA, or + BA or + HxA systems at 298.15 K. Finally, the systems are characterized in terms of the ERAS model, revisiting the previously reported parameters which were determined using volumetric data only [14].

## 2. Experimental

*2.1 Materials*

Information about the purity and source of the pure compounds used along the experiments is collected in Table 1. They were used without further purification. It also shows their densities ($\rho$) at 0.1 MPa and at 298.15 K. These results agree well with literature data.

*2.2 Apparatus and procedure*

Molar quantities were calculated using the relative atomic mass Table of 2015 issued by the Commission on Isotopic Abundances and Atomic Weights (IUPAC) [24].

Densities were obtained using a vibrating-tube densimeter DMA HPM from Anton Paar. The temperature regulation of the densimetric block is insured by the use of a thermostatic bath from Julabo. The standard uncertainty in the temperature is 0.01 K. Experiments were performed at atmospheric pressure, in a static mode. The calibration was carried out using pure octane, dodecane and tridistilled water, and comparing with literature values.

The excess molar enthalpies were determined from heat of mixing measurements performed with a BT2.15 calorimeter from Setaram adapted to work in dynamic mode at constant temperature and pressure. The arrangement is depicted in Figure 1. The fluids flow in stainless steel tubes with an external diameter of 1.6 mm and an internal diameter of 1.0 mm and mix in a custom-made cell. They are injected into the system by means of two syringe pumps model Teledyne ISCO 260 D, which are controlled by a Teledyne ISCO D-Series Pump Controller. Mixtures of different concentrations are obtained varying the volumetric flow rates given by the pumps. These flow rates can be chosen from 1 µL·min$^{-1}$ to 25 mL·min$^{-1}$ with a relative standard uncertainty of 0.5%. The capacity of the pumps is 266.05 mL, and they can be regulated up to a pressure of 52 MPa with a 2% relative standard uncertainty. To ensure the stability of the molar flow rates, the fluids are kept inside the pumps at a constant temperature of 298.15 K by means of a thermostatic bath Fisher Scientific Polystat 36, with a stability of 0.03 K. The relative standard uncertainty in the mole fraction is estimated to be 0.004. The pressure in the system is maintained constant with the help of a pressure regulator located at the end of the flow line, and the pressure relative to the atmospheric pressure is determined by a Keller transducer with a relative standard uncertainty of 0.25% of full scale (40 MPa). For the measurements in this work, the pressure regulator was open to the atmospheric pressure. The temperature of the calorimetric block is regulated by heating a cold can by means of a Setaram G11 Universal Controller. The temperature of the can is maintained constant using a circulating fluid at 10 K below the expected temperature of the experiment, using an external ultra-cryostat Julabo FL1201. The temperature of the block is then regulated using the G11 Universal Controller with a stability of 0.01 K. The temperature of the injected fluids is adjusted to the working temperature with the help of an external precooler and an internal preheater. The external precooler is situated on top of the calorimetric block and is connected in series to the cooler can of the calorimeter and to the ultra-cryostatic bath. The internal preheater is inside the calorimetric block; it supplies the necessary power to reach the exact temperature of the experiment using a heating cartridge, and its temperature is controlled by means of a platinum resistance connected to a Fluke Hart Scientific 2200 PID controller with a stability of 0.01 K. The heat flow is detected by a thermopile, generating an electromotive force (EMF) that is collected by a 6 ½ digit multimeter from Keysight model 34401A and sent to a computer through a GPIB connection. The thermopile EMF, $S$, is converted into the mixing enthalpy through the steady-state relation:

$$H_m^E = \frac{S - S_{BL}}{K(\dot{n}_1 + \dot{n}_2)} \qquad (1)$$

where $K$ is a temperature-dependent calibration constant, $\dot{n}_i$ is the molar flow rate of component $i$ and $S_{BL}$ is the baseline signal, recorded when only one of the fluids is flowing. The constant $K$ is obtained by measuring the $H_m^E$ of the system ethanol + water and comparing the results with reference values from Ott *et al*. [25, 26]. Taking into account uncertainties on fluid flow rates, thermopile calibration $K$, and calorimetric signal noises, the estimated maximum relative standard uncertainty on $H_m^E$ for the set of experimental points in this work is 0.03.

## 3. Results

Data on $H_m^E$ are listed in Table 2. They were fitted to a Redlich-Kister equation [27] by an unweighted linear least-squares regression. The Redlich-Kister equation for the excess property $F^E$ is given by:

$$F^E = x_1(1-x_1)\sum_{i=0}^{k-1} A_i (2x_1 - 1)^i \qquad (2)$$

The number, $k$, of necessary coefficients for this regression has been determined, for each system, by applying an F-test of additional term [28] at 99.5% confidence level. The standard deviations, $\sigma(F^E)$, are defined by:

$$\sigma(F^E) = \left[\frac{1}{N-k}\sum_{j=1}^{N}\left(F_{cal,j}^E - F_{exp,j}^E\right)^2\right]^{1/2} \qquad (3)$$

where the index $j$ takes one value for each of the $N$ data points $F_{\exp,j}^{E}$, and $F_{\text{cal},j}^{E}$ is the corresponding value of the excess property calculated from equation (2).

Excess molar energies of constant volume, $U_{m,V}^{E}$, are given by [29]:

$$U_{m,V}^{E} = H_{m}^{E} - T\frac{\alpha_p}{\kappa_T}V_{m}^{E} \tag{4}$$

where $\alpha_p$ is the isobaric thermal expansion coefficient, $\kappa_T$ is the coefficient of isothermal compressibility and $V_{m}^{E}$ is the excess molar volume. The $U_{m,V}^{E}$ curves of amide + amine systems were obtained at $\Delta x_1 = 0.05$ using smoothed values of $H_{m}^{E}$ and of volumetric properties previously measured [11, 12]. Let us denote by $V_{m,i}$, $\alpha_{p,i}$ and $C_{p,m,i}$ the molar volume, isobaric thermal expansion coefficient and molar isobaric heat capacity of component $i$ respectively, and by $\phi_i = x_i V_{m,i}/(x_1 V_{m,1} + x_2 V_{m,2})$ the volume fraction of component $i$. In the application of equation (4), $\alpha_p$ was assumed ideal ($\alpha_p^{id} = \phi_1 \alpha_{p,1} + \phi_2 \alpha_{p,2}$) for HxA and DBA mixtures; the error in using this assumption is negligible due to the smallness of $V_{m}^{E}$ for these systems and, actually, the difference $U_{m,V}^{E} - H_{m}^{E}$ is not relevant. $\kappa_T$ was obtained from the equation:

$$\kappa_T = \kappa_S + \frac{TV_m \alpha_p^2}{C_{p,m}} \tag{5}$$

with the molar isobaric heat capacity of the mixture, $C_{p,m}$, taken as ideal ($C_{p,m}^{id} = x_1 C_{p,m,1} + x_2 C_{p,m,2}$). The $U_{m,V}^{E}$ curves have also been adjusted to Redlich-Kister polynomials using the same procedure given above.

Table 3 includes the parameters $A_i$ obtained for $F^{E} (= H_{m}^{E}, U_{m,V}^{E})$, together with the standard deviations $\sigma(F^{E})$. Values of $H_{m}^{E}$ at temperature 298.15 K are plotted in Figures 2 and 3, and their corresponding Redlich-Kister regressions in Figures S1 and S2. The corresponding $U_{m,V}^{E}$ curves are depicted in Figures S3 and S4.

## 4. ERAS model

The Extended Real Associated Solution (ERAS) model [15, 30] combines the Real Association Solution Model [31-34] with Flory's thermal equation of state [35-39]. Some important features of this model are now given. (i) The excess molar functions of enthalpy and volume ($F_{m}^{E} = H_{m}^{E}, V_{m}^{E}$) are calculated as the sum of two contributions. The chemical contribution, $F^E_{m,chem}$, arises from hydrogen bonding; the physical contribution, $F^E_{m,phys}$, is related to nonpolar Van der Waals interactions and free volume effects. Expressions for the molar excess functions $H^E_m$ and $V^E_m$ can be found elsewhere [40, 41]. (ii) It is assumed that only consecutive linear association occurs. Accordingly, self-association is described by a chemical equilibrium constant ($K_A$) independent of the chain length of the self-associated species A (in this case, amines), according to the equation:

$$A_m + A \xleftrightarrow{K_A} A_{m+1} \qquad (6)$$

with $m$ ranging from 1 to $\infty$. The cross-association between a self-associated species $A_m$ and a non self-associated compound B (in this study, tertiary amides) is represented by

$$A_m + B \xleftrightarrow{K_{AB}} A_m B \qquad (7)$$

where cross-association constants, $K_{AB}$, are also considered to be independent of the chain length. The molar enthalpies of intermolecular hydrogen-bonding for these two kinds of reactions, $\Delta h^*_A$ and $\Delta h^*_{AB}$, are introduced, and the corresponding equilibrium constants depend on temperature according to them and the Van't Hoff equation. Moreover, negative molar hydrogen-bonding volumes, $\Delta v^*_A$ and $\Delta v^*_{AB}$, are defined in order to take into account the decrease of the core volume of the molecules upon multimer formation. (iii) The $F^E_{m,phys}$ term is derived from the Flory's equation of state [35-39], which is assumed to be valid not only for pure compounds but also for the mixture [42, 43]:

$$\frac{\bar{p}_i \bar{V}_i}{\bar{T}_i} = \frac{\bar{V}_i^{1/3}}{\bar{V}_i^{1/3} - 1} - \frac{1}{\bar{V}_i \bar{T}_i} \qquad (8)$$

where $i$ = A, B or M (mixture). In equation (8), $\bar{V}_i = V_{m,i}/V^*_{m,i}$; $\bar{p}_i = p/p^*_i$; $\bar{T}_i = T/T^*_i$ are the reduced properties for volume, pressure and temperature, respectively. The pure component reduction parameters ($V^*_{m,i}, p^*_i, T^*_i$) are obtained from $p$-$V$-$T$ data (density, isobaric thermal expansion coefficient, and coefficient of isothermal compressibility) and association parameters [42, 43]. The reduction parameters for the mixture $p^*_M$ and $T^*_M$ are calculated from mixing rules [42, 43]. The total relative molecular volumes and surfaces of the compounds were calculated additively on the basis of the group volumes and surfaces recommended by Bondi [44].

## 4.1. Adjustment of ERAS parameters

Values of $V_{m,i}$, $V^*_{m,i}$ and $p^*_i$ of pure compounds [45-47] at $T$ = 298.15 K, needed for calculations, are listed in Table S1 of supplementary material. $K_A$, $\Delta h^*_A$, and $\Delta v^*_A$ of the self-associated amines are known from $H_m^E$ and $V_m^E$ data for the corresponding mixtures with alkanes [45-47], and are also collected in Table S1. The binary parameters to be fitted against $H_m^E$ and $V_m^E$ [11, 12] data of amine + amide systems are then $K_{AB}$, $\Delta h_{AB}^*$, $\Delta v_{AB}^*$ and $X_{AB}$. They are collected in Table 4.

## 5. Discussion

We are referring throughout this section to values of the excess functions and of the thermophysical properties at 298.15 K and at $x_1 = 0.5$, except otherwise specified.

As previously mentioned, DMF and DMA are very polar substances since their dipole moment is 3.7 D [48, 49]. Consequently, their alkane mixtures show immiscibility gaps up to rather high temperatures. Thus, systems formed by DMF and heptane or hexadecane have upper critical solution temperatures (UCST) of 342.55 K [50] and 385.15 K [51] respectively, and the UCST of the DMA + heptane mixture is 309.40 K [52].

Primary and secondary amines are self-associated compounds [30, 45, 46, 53, 54] with lower dipole moments than tertiary amides: 1.3 D (BA) [55], 1.3 D (HxA) [48], 1.0 D (DPA) [55], and 1.1 D (DBA) [55]. For heptane solutions, $H_m^E$/J·mol$^{-1}$ = 1192 (BA) [56], 962 (HxA) [56], 424 (DPA) [57], and 317 (DBA) [57]. We note that $H_m^E$ results are larger for systems with primary amines, and that they decrease with the chain length of the amine. Therefore, these values can be interpreted as arising from the rupture of interactions between like molecules in the mixing process.

Our $H_m^E$ values obtained for amide + amine systems are also positive. We have $H_m^E$(DMF)/J·mol$^{-1}$ = 386 (BA), 660 (HxA), 750 (DPA), 1103 (DBA); and $H_m^E$(DMA)/J·mol$^{-1}$ = 209 (BA), 425 (HxA), 510 (DPA), and 829 (DBA). They can be ascribed to the dominance of contributions from the breaking of amide-amide and amine-amine interactions over that related to the formation of interactions between unlike molecules. Note that $H_m^E$ values of the DMA + cyclohexane mixture are much higher than those of DMA + linear amine systems (Figure S2). The same trend is observed, e.g., when $H_m^E$ results are compared for BA + heptane and *N,N*-dialkylamide systems (Figures S1 and S2). For a fixed amide and along both series of primary or secondary linear amines, $H_m^E$ becomes larger when the chain length of the amine is longer. This suggests that the lower contribution from the breaking of amine-amine interactions in longer amines is overcompensated by the higher contributions which arise from: i) the larger number of amide-amide interactions broken by longer amines; and ii) the lower number and weaker amide-amine interactions created when longer amines are involved, since then the amine group is more sterically hindered.

For a fixed amine, the replacement of DMF by DMA leads to decreased $H_m^E$ values. The difference in size between both amides suggests that the contribution from the disruption of amine-amine interactions should be higher for DMA mixtures. However, the amide group is less sterically hindered in DMF, and we recognize that, in pure state, DMF-DMF interactions are stronger than those between DMA molecules. In fact (see above), UCST(DMF + heptane) > UCST(DMA + heptane). This is also supported by calculations on entropy changes under the action of an electrostatic field and by the application of the Kirkwood-Fröhlich model [14]. Therefore, we can conclude that the breaking of DMF-DMF interactions contributes more positively to $H_m^E$ than the disruption of DMA-DMA interactions, and that the formation of interactions between unlike molecules should contribute more negatively to $H_m^E$ in the case of DMF systems. The mentioned trend suggests that the variation of the contribution of amide-amide interactions is predominant over the other two. The same phenomenon is encountered in 2-alkanone + amine mixtures when the chain length of the 2-alkanone is increased. For example, $H_m^E$(DPA)/J·mol$^{-1}$ = 648 (propanone), 398 (butanone), 281 (2-pentanone) and 161 (2-heptanone) [58].

Interestingly, the replacement of HxA by DPA in systems involving a given amide leads to slightly higher $H_m^E$ values. This can be explained taking into account that, since the amine group is less sterically hindered in HxA, a higher number of interactions between unlike molecules is formed in solutions with this amine and that such interactions are also stronger. It should be noted that the opposite trend is encountered for HxA or DPA + heptane mixtures, and that the difference $H_m^E$(HxA)– $H_m^E$(DPA) for these systems is remarkably higher than that for the corresponding amide solutions: 438 (*n*-heptane); –90 (DMF) and –85 (DMA) (all values in J·mol$^{-1}$). This underlines the relevance of amide-amine interactions in the studied solutions, which had already been mentioned [11, 12]. The previous statement could seem somewhat hasty, since the difference between $H_m^E$ values for amide + HxA or + DPA solutions is rather low. In order to reinforce it, let us remove equation-of-state effects from $H_m^E$ by the calculation of $U_{m,V}^E$ (equation (4)), retaining only interactional contributions. For our mixtures, $U_{m,V}^E$/J·mol$^{-1}$ = 489 (DMF + BA), 667 (DMF + HxA), 852 (DMF + DPA), and 1096 (DMF + DBA); 283 (DMA + BA), 423 (DMA + HxA), 590 (DMA + DPA), and 809 (DMA + DBA). The difference between $U_{m,V}^E$ values of amide + HxA or + DPA solutions is approximately twice the corresponding difference between their $H_m^E$ results. This supports our previous discussion on the importance of amide-amine interactions. Eventually, let us point out the large and negative value of the $H_m^E$ of the system *N*-methylacetamide + HxA (–1000 J·mol$^{-1}$, $T$ = 363.15 K) [23], for which the formation of amide-amine interactions is dominant by far.

The excess molar volumes, $V_m^E$/cm$^3$·mol$^{-1}$, of the considered mixtures are either negative or small and positive [11, 12]: –0.263 (DMF + BA), –0.021 (DMF + HxA), –0.289 (DMF + DPA), and 0.018 (DMF+DBA); –0.194 (DMA + BA), 0.006 (DMA + HxA), –0.228 (DMA + DPA), and 0.055 (DMA + DBA). It is to be noted that $H_m^E$ and $V_m^E$ change in line, which reveals that the interactional contribution to $V_m^E$ is relevant. However, positive $H_m^E$ values together with negative $V_m^E$ results are indicative of the existence of structural effects [59]. Similar structural effects are also encountered in amine + *n*-alkane systems; for example, see the low value of $V_m^E$ / cm$^3$·mol$^{-1}$ in DBA + heptane, 0.0675 (DBA) [60], and the negative one of the DBA + hexane system, –0.1854 cm$^3$·mol$^{-1}$ [61].

Mixtures of DMF or DMA with aniline contrast drastically with those of linear primary or secondary amines. The dipole moment of aniline (1.51 D [49]) is higher than that of linear primary and secondary amines, and proximity effects between the phenyl ring and the amine group lead to strong dipolar interactions between aniline molecules. As a consequence, aniline + *n*-alkane mixtures are characterized by relatively high UCST (343.11 K for the heptane solution [62]). When aniline molecules are mixed with DMF or DMA molecules, very strong interactions between unlike molecules are created, and we have $H_m^E$/J·mol$^{-1}$ = − 2946 (DMF + aniline) [21]; − 352 (DMA + aniline) [22]. Similarly, large differences are also encountered between values of the excess relative permittivity for the DMF + linear primary or secondary amine or + aniline mixtures [13, 14]: –0.864 (DMF + BA), –1.262 (DMF + HxA), –1.372 (DMF + DPA), –1.733 (DMF + DBA), 1.806 (DMF + aniline). It must be observed that $H_m^E$ values are very different for DMF and DMA + aniline systems, newly remarking that interactions between unlike molecules are much more relevant in DMF systems. The rather large and negative $V_m^E$ / cm$^3$·mol$^{-1}$ results for the mentioned aniline solutions ( –0.6615 (DMF + aniline) [63] and –0.6092 (DMA + aniline, $T$ = 303.15 K) [64]) are in agreement with the $H_m^E$ values and underline the importance of the interactional contribution to $V_m^E$.

### 5.1. ERAS results

Results from ERAS are collected in Tables 5 and 6 and are shown graphically in Figures 2-5. Both excess functions, $H_m^E$ and $V_m^E$, are reasonably well represented by the model. Larger differences for $V_m^E$ results are encountered for mixtures characterized by low $V_m^E$ values, as then the overall result is obtained from the difference of two large magnitudes of different sign: the positive physical contribution and the negative chemical contribution (Table 6). ERAS calculations indicate that a better agreement with $V_m^E$ data is obtained when the chemical contribution to $V_m^E$ is higher (in absolute value). This occurs for the BA or DPA + DMF systems (Figure 4). In terms of the model, such excess molar volumes are mainly determined by the interactions between unlike molecules. In contrast, structural effects seem to be relatively more important in the BA or DPA + DMA mixtures (Table 6). On the other hand, we note that ERAS results on $H_m^E$ are, as an average, better for DMA systems (Table 5). This suggests that, in such a case, physical interactions are more properly described by ERAS, that is, dipolar interactions are more relevant in DMF mixtures, particularly in the BA solution.

The low $K_{AB}$ and $|\Delta h_{AB}^*|$ values (Table 4) indicate that solvation effects are not relevant and that the enthalpy of the H bonds between unlike molecules is weak. The large $X_{AB}$ values (Table 4) reveal that the physical contribution is important, particularly with regards to $H_m^E$. The present ERAS parameters largely differ from those determined for 1-alkanol + linear primary or secondary amine systems, which are characterized by strong solvation effects and, in consequence, by large $K_{AB}$ and $\Delta h_{AB}^*$ values and low $X_{AB}$ values. For example, for the 1-hexanol + HxA mixture at 298.15 K: $K_{AB}$ = 800; $\Delta h_{AB}^*$ = –36 kJ·mol$^{-1}$; $X_{AB}$ = 5 J·cm$^{-3}$ [47]. As we have pointed out (see above), aniline-amide interactions are rather strong and, accordingly, the corresponding ERAS parameters are also very different. We have: $K_{AB}$ = 70 (DMF); 2.2 (DMA); $\Delta h_{AB}^*$/ kJ·mol$^{-1}$ = –22 (DMF; DMA); $\Delta v_{AB}^*$/cm$^3$·mol$^{-1}$ = –11.1 (DMF); –20 (DMA); $X_{AB}$/ J·cm$^{-3}$ = 4 (DMF); 3.2 (DMA) [14].

## 6. Conclusion

Excess molar enthalpies of amide (DMF or DMA) + linear primary or secondary amine (BA, HxA, DPA or DBA) have been reported at $T$ = 298.15 K and $p$ = 0.1 MPa. The positive $H_m^E$ values arise from the dominant contribution from the rupture of amide-amide and amine-amine interactions along mixing. Dipolar interactions are stronger in DMF systems. DMA mixtures show lower $H_m^E$ values for a fixed amine, suggesting that the variation of the rupture of amide-amide interactions is the predominant effect. Results on $H_m^E$ and $U_{m,V}^E$ reveal that interactions between unlike molecules are stronger in mixtures containing HxA compared to those with DPA for a given amide. Negative or small positive $V_m^E$ values point to the existence of important structural effects in the investigated solutions. The binary interaction parameters of the ERAS model have been adjusted to fit $H_m^E$ and $V_m^E$ curves simultaneously, and these properties are represented with a rather good degree of approximation. The results from the model suggest that physical interactions are important when calculating the excess functions of the mixtures under study.

## Acknowledgements

F. Hevia is grateful to J.-Y. Coxam and K. Ballerat-Busserolles for the opportunity to do the experimental part of this work at their laboratory at *Institut de Chimie de Clermont-Ferrand*, and also acknowledges *Ministerio de Educación, Cultura y Deporte* for the grant FPU14/04104 and for the complementary grants EST16/00824 and EST17/00292. In addition, the authors FH, JAG, IGF and JCC gratefully acknowledge the financial support received from the Consejería de Educación de Castilla y León, under Project VA100G19 (Apoyo a GIR, BDNS: 425389).

Table 1

Description, source and purity of the pure liquids and their density, $\rho$, at temperature $T = 298.15$ K and pressure $p = 0.1$ MPa.[b]

| Chemical name | CAS Number | Source | Purity[a] | $\rho$ / g·cm$^{-3}$ Exp. | $\rho$ / g·cm$^{-3}$ Lit. |
|---|---|---|---|---|---|
| N,N-dimethylformamide (DMF) | 68-12-2 | Sigma-Aldrich | 0.9996 | 0.94378 | 0.944163 [65] |
| N,N-dimethylacetamide (DMA) | 127-19-5 | Honeywell | >0.999 | 0.93614 | 0.936233 [66] |
| N-propylpropan-1-amine (DPA) | 142-84-7 | Aldrich | 0.999 | 0.73337 | 0.73321 [67] |
| N-butylbutan-1-amine (DBA) | 111-92-2 | Aldrich | 0.997 | 0.75570 | 0.755457 [68] |
| butan-1-amine (BA) | 109-73-9 | Sigma-Aldrich | 0.9978 | 0.73218 | 0.73233 [69] |
| hexan-1-amine (HxA) | 111-26-2 | Aldrich | 0.999 | 0.76016 | 0.76013 [70] |

[a] In mole fraction. By gas chromatography. Provided by the supplier.

[b] The standard uncertainties are: $u(T) = 0.01$ K, $u(p) = 1$ kPa. The relative standard uncertainty is: $u_r(\rho) = 0.0012$.

Table 2

Excess molar enthalpies, $H_m^E$, of amide (1) + amine (2) liquid mixtures as functions of the mole fraction of the amide, $x_1$, at temperature $T = 298.15$ K and pressure $p = 0.1$ MPa. [a]

| $x_1$ | $H_m^E$ / J·mol$^{-1}$ | $x_1$ | $H_m^E$ / J·mol$^{-1}$ | $x_1$ | $H_m^E$ / J·mol$^{-1}$ | $x_1$ | $H_m^E$ / J·mol$^{-1}$ |
|---|---|---|---|---|---|---|---|
| | | | DMF (1) + DPA (2) | | | | |
| 0.0358 | 88 | 0.3016 | 595 | 0.5505 | 753 | 0.7983 | 542 |
| 0.1002 | 241 | 0.3483 | 657 | 0.6012 | 743 | 0.8485 | 441 |
| 0.1512 | 351 | 0.4005 | 695 | 0.6587 | 708 | 0.8991 | 318 |
| 0.1984 | 446 | 0.4495 | 737 | 0.6997 | 673 | 0.9504 | 170 |
| 0.2504 | 529 | 0.5005 | 748 | 0.7518 | 613 | | |
| | | | DMF (1) + DBA (2) | | | | |
| 0.0491 | 188 | 0.3006 | 877 | 0.5505 | 1118 | 0.7994 | 815 |
| 0.1014 | 386 | 0.3510 | 956 | 0.5990 | 1098 | 0.8503 | 669 |
| 0.1488 | 543 | 0.4021 | 1032 | 0.6500 | 1060 | 0.9000 | 488 |
| 0.2006 | 680 | 0.4497 | 1082 | 0.6997 | 1001 | 0.9501 | 270 |
| 0.2482 | 782 | 0.4982 | 1113 | 0.7506 | 921 | | |
| | | | DMF (1) + BA (2) | | | | |
| 0.0510 | 89 | 0.2498 | 320 | 0.5008 | 384 | 0.7559 | 245 |
| 0.1009 | 166 | 0.3007 | 352 | 0.5463 | 374 | 0.8005 | 208 |
| 0.1496 | 226 | 0.3496 | 373 | 0.6008 | 355 | 0.8495 | 158 |
| 0.1703 | 247 | 0.4008 | 391 | 0.6463 | 326 | 0.9003 | 106 |
| 0.2005 | 279 | 0.4508 | 393 | 0.7011 | 289 | 0.9519 | 55 |
| | | | DMF (1) + HxA (2) | | | | |
| 0.0514 | 118 | 0.3006 | 533 | 0.5505 | 659 | 0.8005 | 441 |
| 0.1007 | 219 | 0.3519 | 586 | 0.5982 | 648 | 0.8506 | 354 |
| 0.1516 | 318 | 0.4007 | 619 | 0.6507 | 615 | 0.8995 | 246 |
| 0.2009 | 394 | 0.4518 | 648 | 0.7002 | 569 | 0.9502 | 129 |
| 0.2522 | 461 | 0.5003 | 661 | 0.7503 | 510 | | |
| | | | DMA (1) + DPA (2) | | | | |
| 0.0597 | 98 | 0.3009 | 399 | 0.5497 | 519 | 0.8009 | 385 |
| 0.0996 | 166 | 0.3462 | 435 | 0.5972 | 509 | 0.8504 | 318 |
| 0.1496 | 235 | 0.4062 | 477 | 0.6495 | 493 | 0.8989 | 239 |
| 0.2011 | 300 | 0.4491 | 495 | 0.6975 | 473 | 0.9368 | 156 |
| 0.2497 | 358 | 0.5035 | 514 | 0.7478 | 437 | 0.9507 | 126 |
| | | | DMA (1) + DBA (2) | | | | |
| 0.0499 | 150 | 0.3002 | 673 | 0.5507 | 832 | 0.8004 | 618 |
| 0.0976 | 276 | 0.3526 | 732 | 0.6007 | 825 | 0.8512 | 508 |
| 0.1510 | 405 | 0.3983 | 774 | 0.6477 | 803 | 0.9019 | 371 |
| 0.1969 | 503 | 0.4472 | 807 | 0.6968 | 756 | 0.9507 | 203 |
| 0.2472 | 590 | 0.5007 | 833 | 0.7481 | 698 | | |
| | | | DMA (1) + BA (2) | | | | |
| 0.0498 | 48 | 0.3004 | 189 | 0.5509 | 201 | 0.9002 | 64 |
| 0.0992 | 89 | 0.3493 | 202 | 0.6005 | 195 | 0.9594 | 27 |
| 0.1518 | 123 | 0.4015 | 207 | 0.7008 | 160 | | |
| 0.1985 | 152 | 0.4508 | 209 | 0.7995 | 120 | | |
| 0.2512 | 165 | 0.5005 | 208 | 0.8502 | 94 | | |
| | | | DMA (1) + HxA (2) | | | | |
| 0.0509 | 71 | 0.3018 | 338 | 0.5493 | 424 | 0.7996 | 296 |
| 0.1005 | 141 | 0.3484 | 369 | 0.6002 | 416 | 0.8513 | 238 |
| 0.1507 | 196 | 0.3980 | 402 | 0.6492 | 401 | 0.9008 | 171 |
| 0.1994 | 244 | 0.4502 | 419 | 0.7001 | 375 | 0.9503 | 91 |
| 0.2488 | 295 | 0.5006 | 423 | 0.7491 | 341 | | |

[a] The standard uncertainties are: $u(T)$ = 0.01 K, $u(p)$ = 1 kPa. The relative standard uncertainty is: $u_r(x_1)$ = 0.004. The relative combined expanded uncertainty (0.95 level of confidence) is $U_{rc}(H_m^E) = 0.06$.

Table 3

Coefficients $A_i$ and standard deviations, $\sigma(F^E)$ (equation (3)), for the representation of $F^E$ at temperature $T = 298.15$ K and pressure $p = 0.1$ MPa for amide + amine liquid mixtures by equation (2).

| Property $F^E$ | System | $A_0$ | $A_1$ | $A_2$ | $A_3$ | $\sigma(F^E)$ |
|---|---|---|---|---|---|---|
| $H_m^E$ / J·mol$^{-1}$ | DMF + DPA | 2999 | 476 | 178 | | 4 |
| | DMF + DBA | 4410 | 731 | 634 | | 7 |
| | DMF + BA | 1545 | −374 | −75 | | 1.9 |
| | DMF + HxA | 2639 | 238 | −93 | | 4 |
| | DMA + DPA | 2038 | 444 | 274 | | 4 |
| | DMA + DBA | 3314 | 476 | 544 | 255 | 2 |
| | DMA + BA | 834 | −158 | | | 3 |
| | DMA + HxA | 1699 | 234 | | | 3 |
| $U_{m,V}^E$ / J·mol$^{-1}$ | DMF + DPA | 3408 | 657 | 338 | | 0.7 |
| | DMF + DBA | 4385.4 | 841 | 758 | | 0.5 |
| | DMF + BA | 1954.3 | −193 | 45 | | 0.4 |
| | DMF + HxA | 2669.7 | 376 | | | 0.7 |
| | DMA + DPA | 2359.7 | 457 | 368 | | 0.4 |
| | DMA + DBA | 3237.3 | 436 | 532 | 250 | 0.3 |
| | DMA + BA | 1131 | −59 | 93 | | 0.3 |
| | DMA + HxA | 1690 | 269 | 31 | | 0.6 |

Table 4

ERAS parameters for amine (A) + DMF (B) or + DMA (B) liquid mixtures at temperature 298.15 K and pressure 0.1 MPa. $K_{AB}$, association constant of component A with component B; $\Delta h^*_{AB}$, association enthalpy of component A with component B; $\Delta v^*_{AB}$, association volume of component A with component B; $X_{AB}$, physical parameter.

| System | $K_{AB}$ | $\Delta h^*_{AB}$ / kJ·mol$^{-1}$ | $\Delta v^*_{AB}$ / cm$^3$·mol$^{-1}$ | $X_{AB}$ / J·cm$^{-3}$ |
|---|---|---|---|---|
| BA + DMF | 1.3 | –9 | –3.5 | 24.5 |
| HxA + DMF | 1 | –9 | –4.6 | 36.0 |
| BA + DMA | 1.3 | –9 | –2.5 | 13.4 |
| HxA + DMA | 1 | –9 | –2.8 | 23.3 |
| DPA + DMF | 1 | –2 | –2.5 | 23.8 |
| DBA + DMF | 1 | –2 | –3.8 | 47.8 |
| DPA + DMA | 1 | –2 | –1.2 | 15.0 |
| DBA + DMA | 1 | –2 | –2.2 | 35.1 |

Table 5

Excess molar enthalpies ($H_m^E$) at equimolar composition, temperature 298.15 K and pressure 0.1 MPa, of amine (A) + DMF (B) or DMA (B) liquid mixtures, and standard deviations, $\sigma(H_m^E)$.

| System | N | $H_m^E$ / J·mol$^{-1}$ | | $\sigma(H_m^E)$ / J·mol$^{-1}$ | |
|---|---|---|---|---|---|
| | | Exp. | ERAS | Exp.[a] | ERAS[b] |
| BA + DMF | 20 | 386 | 380 | 1.9 | 50 |
| HxA + DMF | 19 | 660 | 651 | 4 | 51 |
| BA + DMA | 17 | 744 | 740 | 4 | 22 |
| HxA + DMA | 19 | 1102 | 1137 | 7 | 79 |
| DPA + DMF | 19 | 208 | 211 | 3 | 16 |
| DBA + DMF | 19 | 425 | 407 | 3 | 16 |
| DPA + DMA | 20 | 509 | 521 | 4 | 13 |
| DBA + DMA | 19 | 828 | 842 | 2 | 32 |

[a] Obtained from equation (3).

[b] Defined as: $\sigma(H_m^E) = \left[\dfrac{1}{N}\sum_{j=1}^{N}\left(F_{\text{ERAS},j}^E - F_{\exp,j}^E\right)^2\right]^{1/2}$, with notation similar to equation (3).

Table 6

Excess molar volumes ($V_\text{m}^\text{E}$) at equimolar composition, temperature 298.15 K and pressure 0.1 MPa, of amine (A) + DMF (B) or DMA (B) liquid mixtures[a]. The chemical, $V_{\text{m,chem}}^\text{E}$, and physical, $V_{\text{m,phys}}^\text{E}$, contributions to this excess function within ERAS model are also listed.

| System | $V_\text{m}^\text{E}$ / cm$^3$·mol$^{-1}$ | | | |
|---|---|---|---|---|
| | Exp. | $V_{\text{m,chem}}^\text{E}$ | $V_{\text{m,phys}}^\text{E}$ | ERAS |
| BA + DMF  | −0.263 | −0.330 | 0.071  | −0.259 |
| HxA + DMF | −0.021 | −0.501 | 0.478  | 0.023  |
| DPA + DMF | −0.289 | −0.223 | −0.069 | −0.292 |
| DBA + DMF | 0.018  | −0.628 | 0.648  | 0.020  |
| BA + DMA  | −0.194 | −0.122 | −0.081 | −0.203 |
| HxA + DMA | 0.006  | 0.230  | −0.189 | 0.041  |
| DPA + DMA | −0.228 | 0.022  | −0.257 | −0.235 |
| DBA + DMA | 0.055  | −0.303 | 0.349  | 0.046  |

[a] Source of experimental data: [11] for DMF mixtures, [12] for DMA systems.

Figure 1

Schematic view of the experimental setup used to determine excess molar enthalpies.

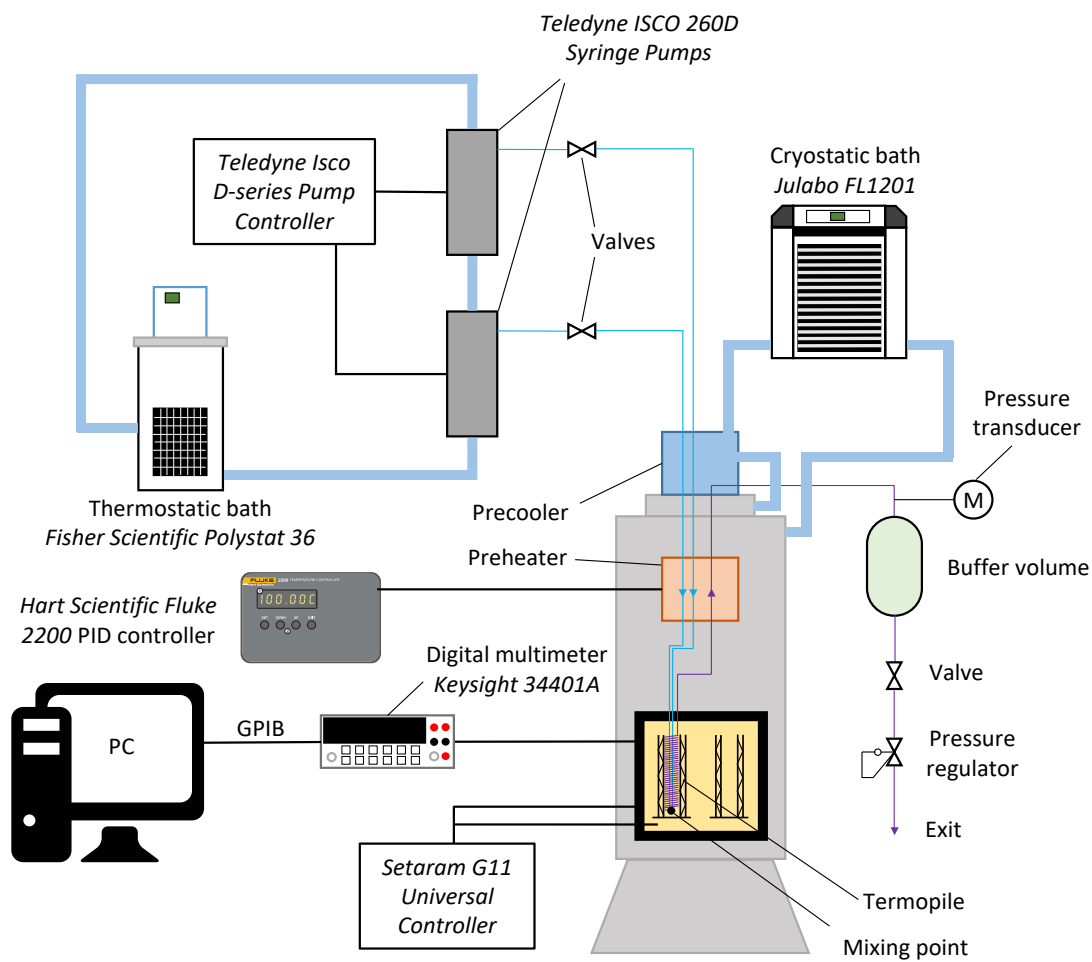

Figure 2

Excess molar enthalpies, $H_m^E$, of DMF (1) + amine (2) liquid mixtures at 0.1 MPa and 298.15 K. Full symbols, experimental values (this work): (●), BA; (■), HxA; (▲), DPA; (♦), DBA. Solid lines, ERAS results using interaction parameters listed in Table 4.

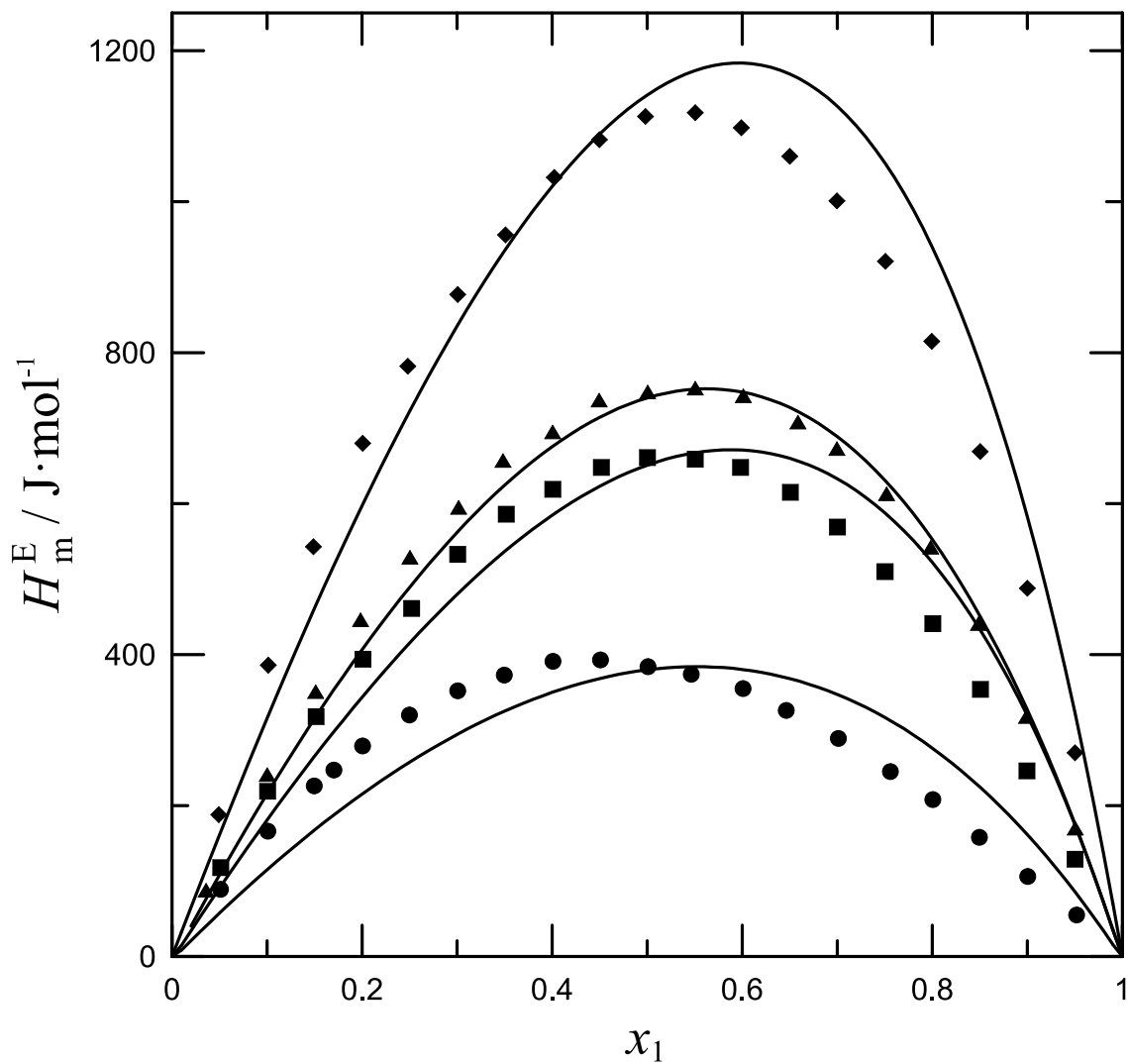

Figure 3

Excess molar enthalpies, $H_m^E$, of DMA (1) + amine (2) liquid mixtures at 0.1 MPa and 298.15 K. Full symbols, experimental values (this work): (●), BA; (■), HxA; (▲), DPA; (♦), DBA. Solid lines, ERAS results using interaction parameters listed in Table 4.

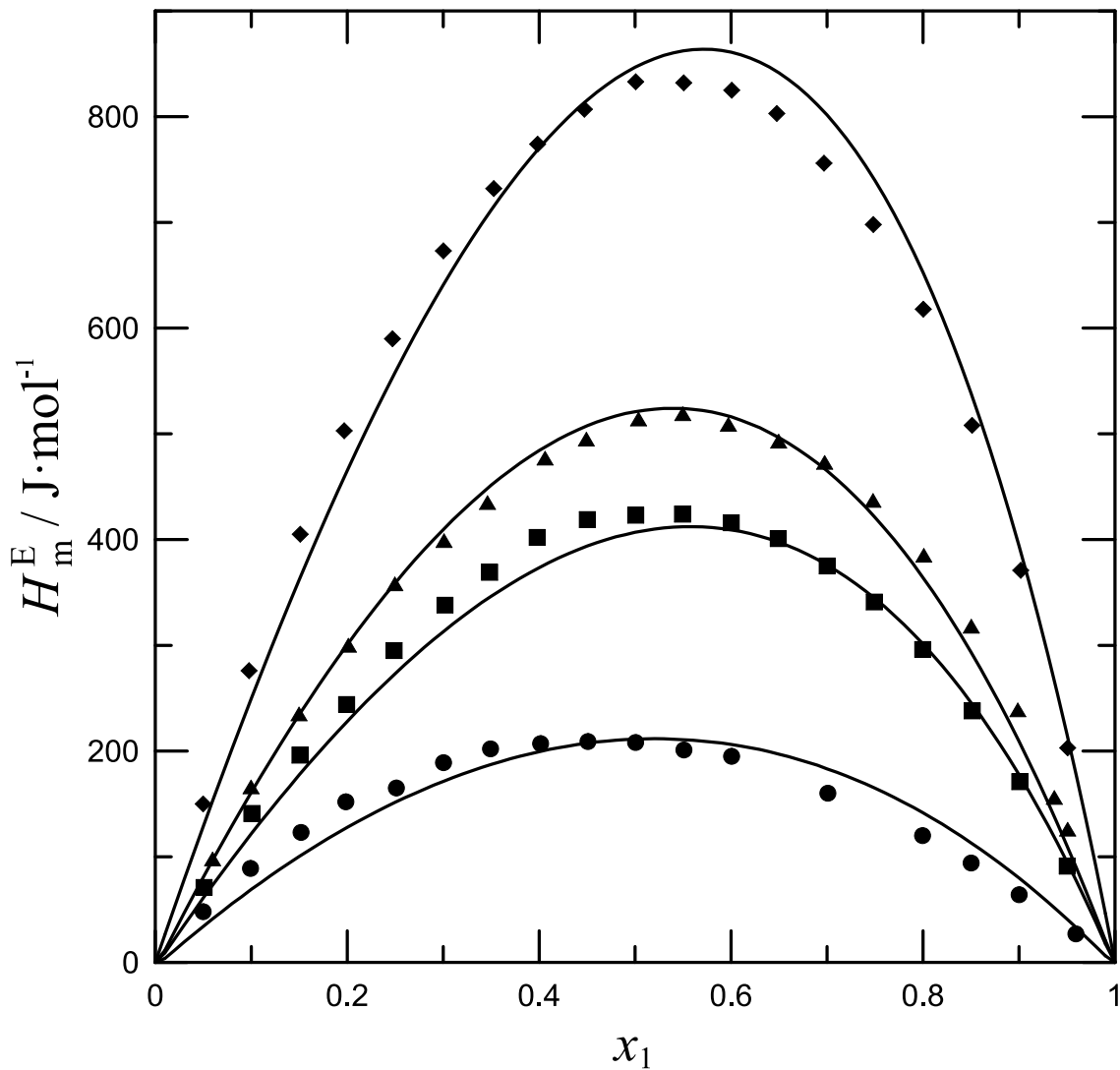

Figure 4

Excess molar volumes, $V_m^E$, of DMF (1) + amine (2) liquid mixtures at 0.1 MPa and 298.15 K. Full symbols, experimental values [11]: (●), BA; (■), HxA; (▲), DPA; (♦), DBA. Solid lines, ERAS results using interaction parameters listed in Table 4.

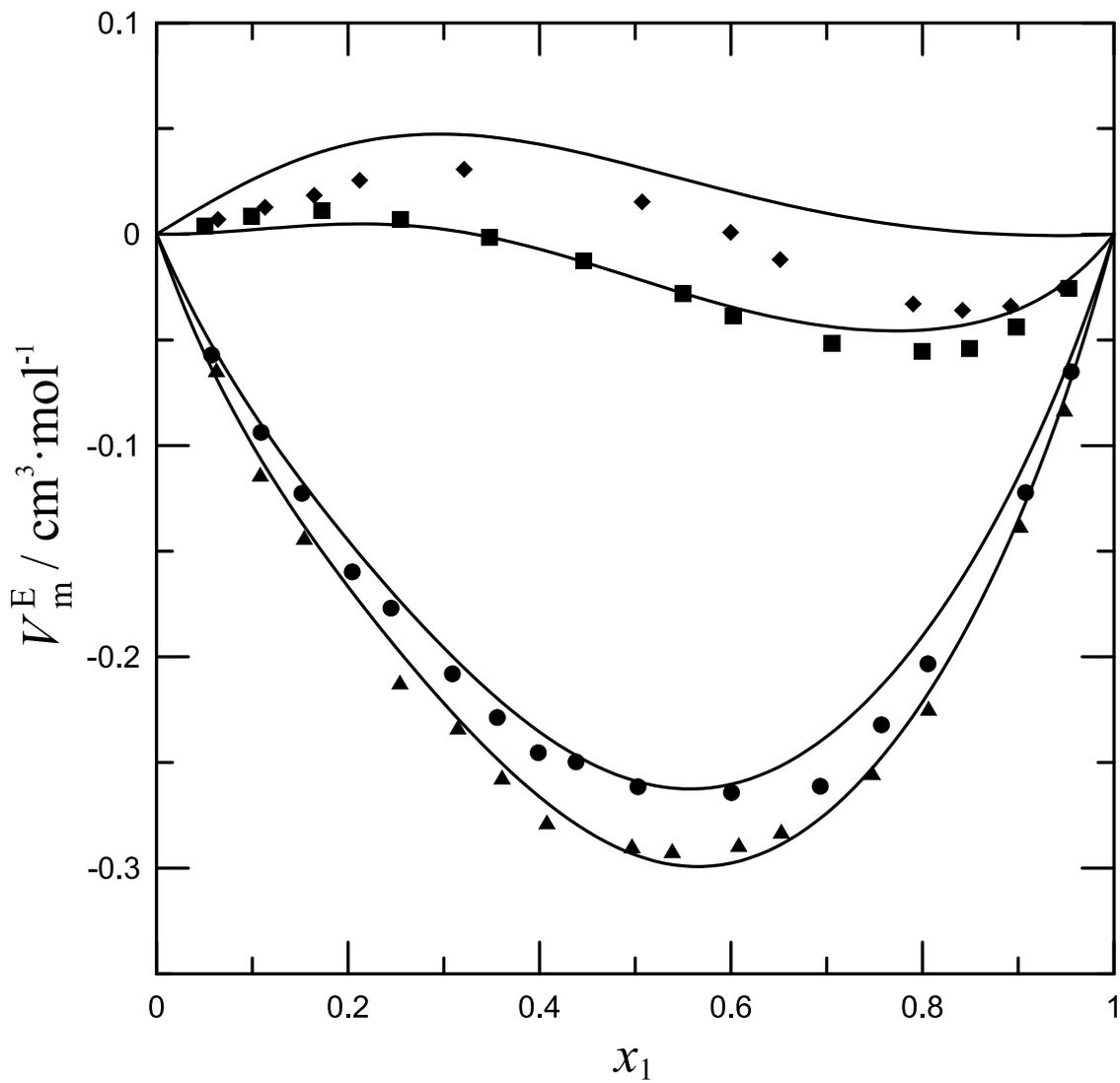

Figure 5

Excess molar volumes, $V_m^E$, of DMA (1) + amine (2) liquid mixtures at 0.1 MPa and 298.15 K. Full symbols, experimental values [12]: (●), BA; (■), HxA; (▲), DPA; (♦), DBA. Solid lines, ERAS results using interaction parameters listed in Table 4.

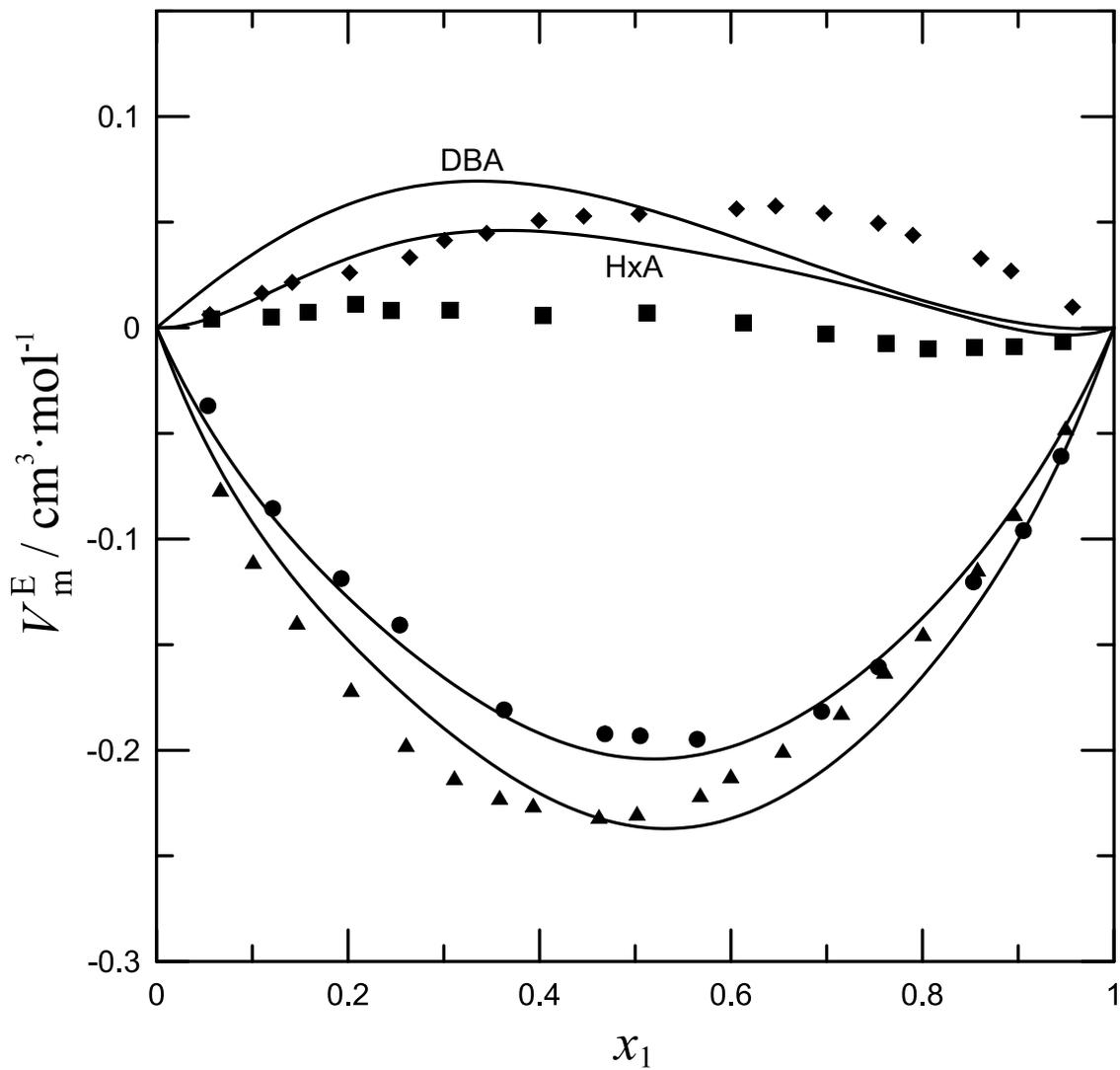

# Thermodynamics of amide + amine mixtures. 5. Excess molar enthalpies of *N,N*-dimethylformamide or *N,N*-dimethylacetamide + *N*-propylpropan-1-amine, + *N*-butylbutan-1-amine, + butan-1-amine, or + hexan-1-amine systems at 298.15 K. Application of the ERAS model

## Supplementary material

Fernando Hevia[1], Karine Ballerat-Busserolles[2], Yohann Coulier[2], Jean-Yves Coxam[2], Juan Antonio González[1]*, Isaías García de la Fuente[1], José Carlos Cobos[1]

[1] G.E.T.E.F., Departamento de Física Aplicada, Facultad de Ciencias, Universidad de Valladolid. Paseo de Belén, 7, 47011 Valladolid, Spain.

[2] Institut de Chimie de Clermont Ferrand, University Clermont Auvergne, CNRS UMR 6296, SIGMA Clermont, F-63000 Clermont-Ferrand, France.

*e-mail: jagl@termo.uva.es; Tel: +34-983-423757

Table S1

ERAS parameters[a] for pure liquids at temperature 298.15 K and pressure 0.1 MPa.

| Compound | $V_{m,i}$ / cm$^3$·mol$^{-1}$ | $K_i$ | $\Delta h_i^*$ / kJ·mol$^{-1}$ | $\Delta v_i^*$ / cm$^3$·mol$^{-1}$ | $V_{m,i}^*$ / cm$^3$·mol$^{-1}$ | $p_i^*$ / J·cm$^{-3}$ |
|---|---|---|---|---|---|---|
| BA[b]  | 99.89  | 0.96 | –13.2 | –2.8 | 77.59  | 565.7 |
| HxA[b] | 133.11 | 0.78 | –13.2 | –2.8 | 106.87 | 495.0 |
| DPA[c] | 138.07 | 0.55 | –7.5  | –2.8 | 106.50 | 526.0 |
| DBA[c] | 171.03 | 0.16 | –4.5  | –2.8 | 135.86 | 466.2 |
| DMF[d] | 77.44  | 0    | 0     | 0    | 62.07  | 714.1 |
| DMA[d] | 93.04  | 0    | 0     | 0    | 75.56  | 649.5 |

[a] $V_{m,i}$, molar volume; $K_i$, equilibrium constant; $V_{m,i}^*$ and $p_i^*$, reduction parameters for volume and pressure, respectively; $\Delta h_i^*$, hydrogen-bonding enthalpy; $\Delta v_i^*$, self-association volume; [b] Ref. [1]; [c] Ref. [2]; [d] [3].

Figure S1

Excess molar enthalpies, $H_m^E$, of DMF (1) + amine (2), or BA (1) + heptane (2) liquid mixtures at 0.1 MPa and 298.15 K. Full symbols, DMF (1) + amine (2) experimental values (this work): (●), BA; (■), HxA; (▲), DPA; (♦), DBA. Solid lines, calculations with equation (2) using the coefficients from Table 3. Dashed line, BA (1) + heptane (2) [4].

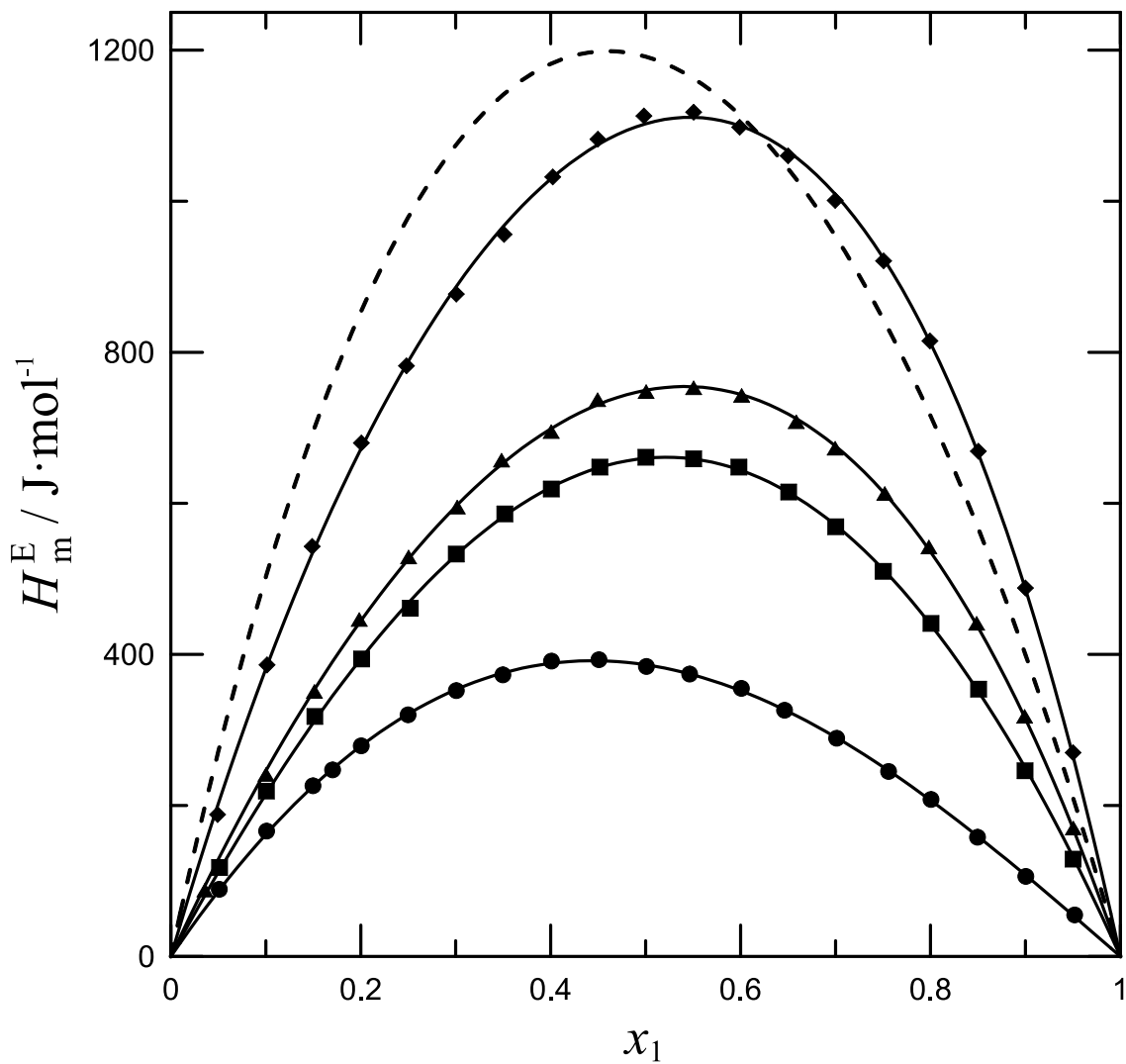

Figure S2

Excess molar enthalpies, $H_m^E$, of DMA (1) + amine (2), or + cyclohexane (2) liquid mixtures at 0.1 MPa and 298.15 K. Full symbols, DMA (1) + amine (2) experimental values (this work): (●), BA; (■), HxA; (▲), DPA; (♦), DBA. Solid lines, calculations with equation (2) using the coefficients from Table 3. Dashed line, DMA (1) + cyclohexane (2) [5].

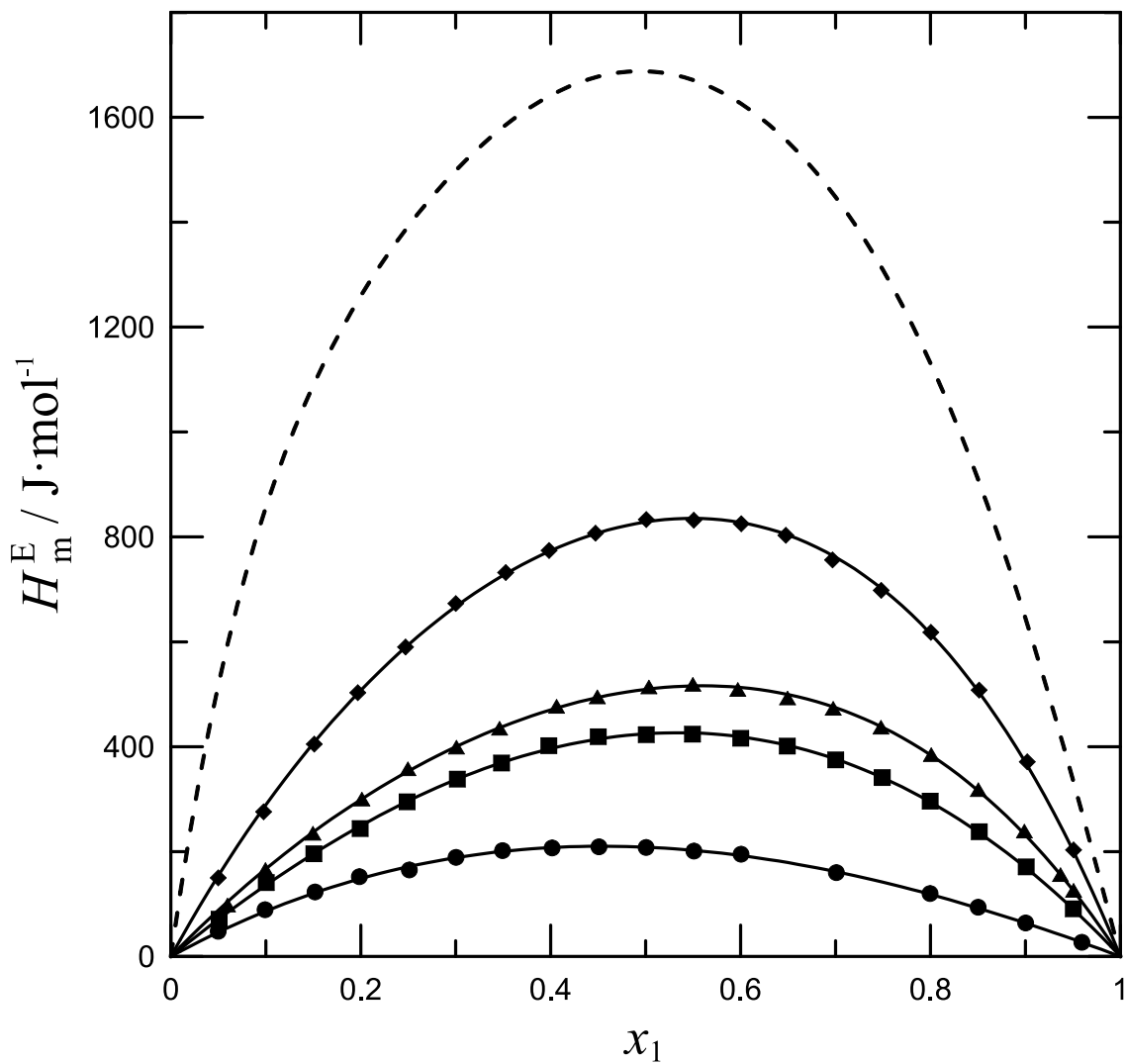

Figure S3

Excess molar energies at constant volume, $U_{m,V}^{E}$, of DMF (1) + amine (2) liquid mixtures at 0.1 MPa and 298.15 K. Full symbols, calculations at $\Delta x_1 = 0.05$ from smoothed experimental values: (●), BA; (■), HxA; (▲), DPA; (♦), DBA. Solid lines, calculations with equation (2) using the coefficients from Table 3.

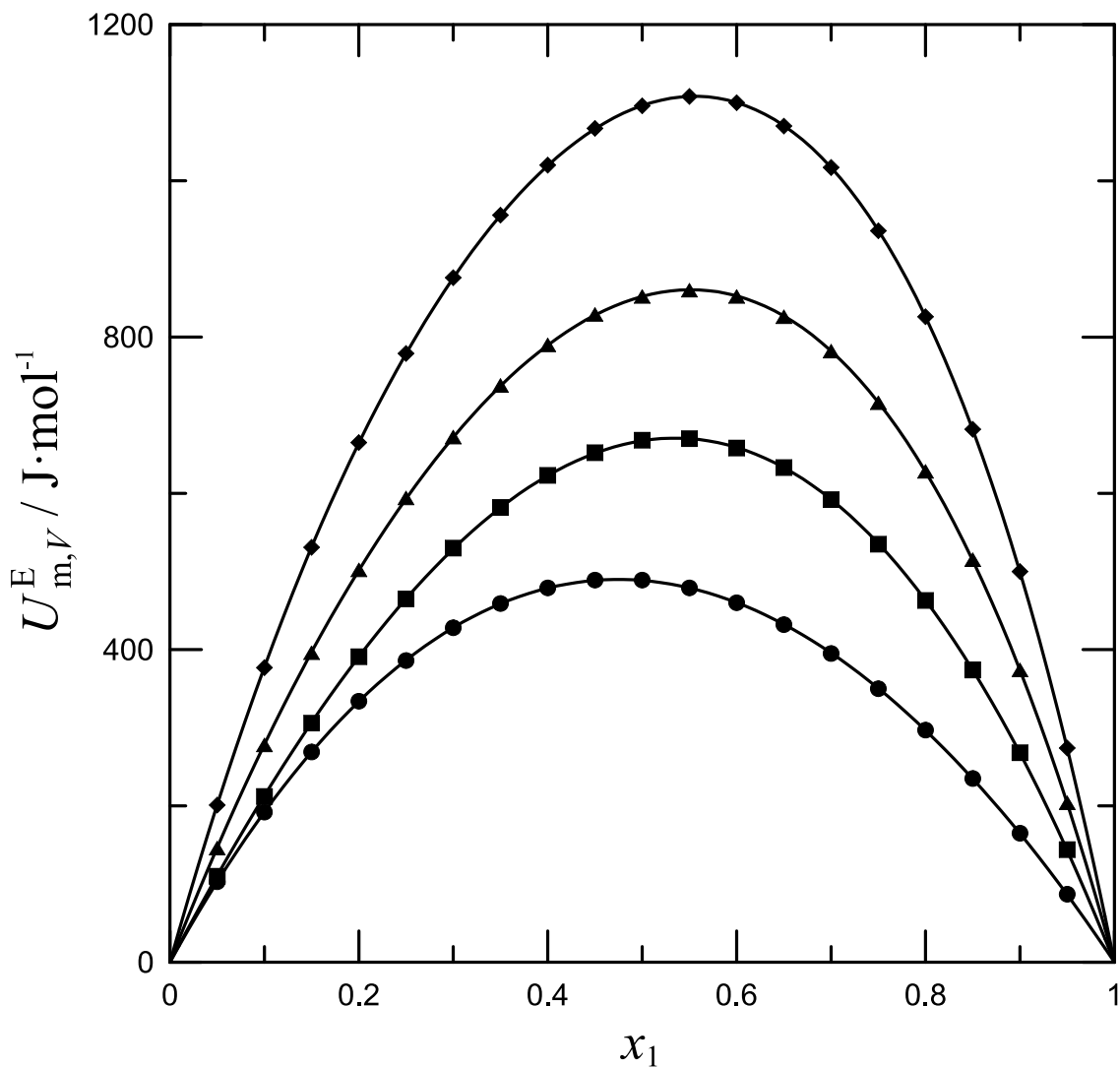

Figure S4

Excess molar energies at constant volume, $U_{m,V}^{E}$, of DMA (1) + amine (2) liquid mixtures at 0.1 MPa and 298.15 K. Full symbols, calculations at $\Delta x_1 = 0.05$ from smoothed experimental values: (●), BA; (■), HxA; (▲), DPA; (♦), DBA. Solid lines, calculations with equation (2) using the coefficients from Table 3.

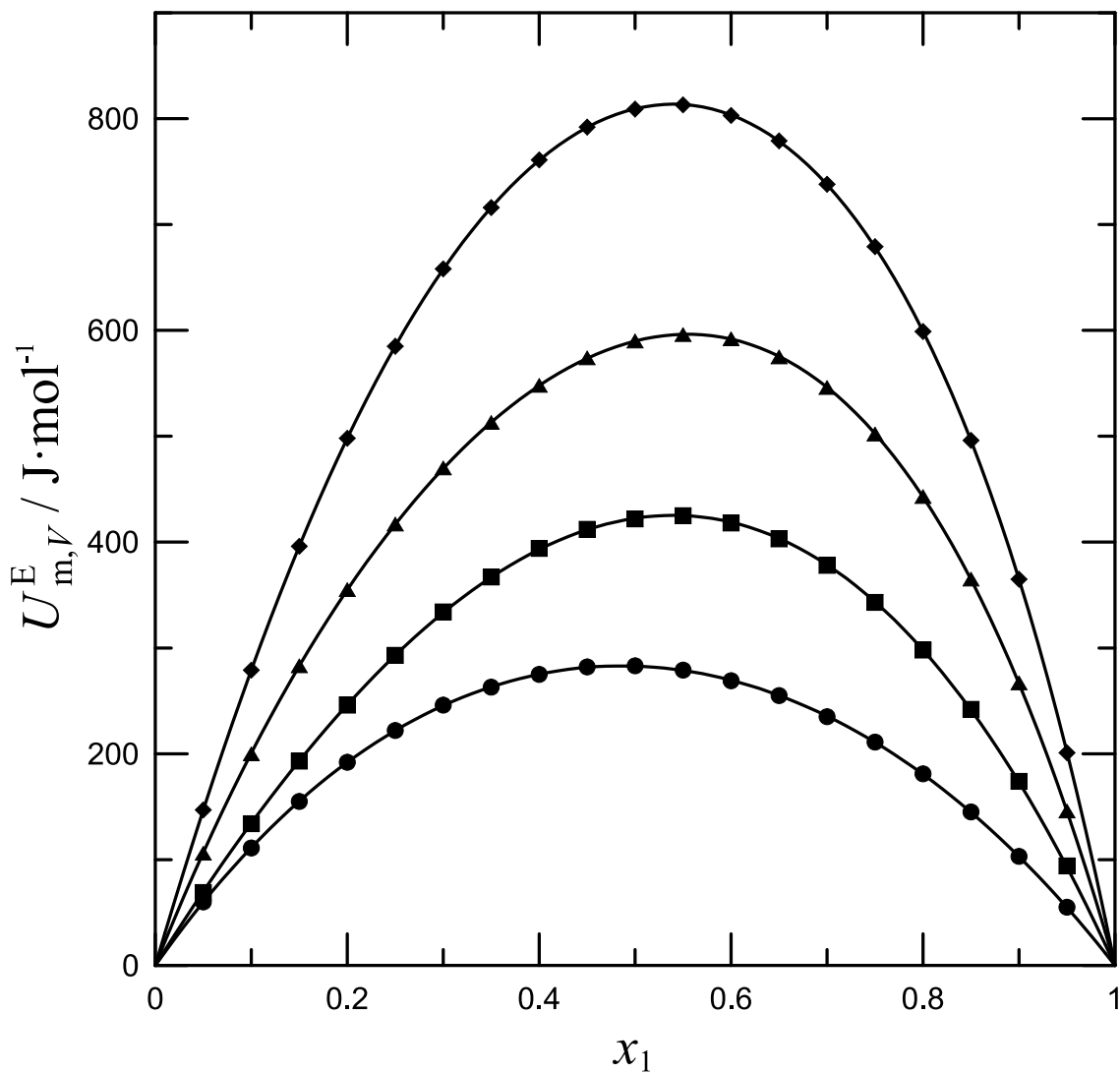

# References for supplementary material